\documentclass[smallextended]{svjour3}      
\smartqed %
\usepackage{graphicx}%
\usepackage[numbers]{natbib}
\begin{document}
\title{Spacetime Perspective of Gravitational Lensing in Perturbed Cosmologies}


\author{Thomas P. Kling       \and
       Sophia MacQueen Pooler 
}

\institute{Thomas P. Kling (corresponding author)
           \at
              Dept. of Physics, Photonics and Optical Engineering, Bridgewater State University,
Bridgewater, MA 02325 \\
              Tel.: +1-508-531-2895\\
              Fax: +1-508-531-1785\\
              \email{tkling@bridgew.edu}    
              \and
            Sophia MacQueen Pooler \at
              Dept. of Physics, Photonics and Optical Engineering, Bridgewater State University,
Bridgewater, MA 02325 \\
}
\date{Received: date / Accepted: date}
\maketitle

\begin{abstract}

\noindent We develop the spacetime approach to gravitational lensing by spherically symmetric perturbations of flat, cosmological constant-dominated Friedman-Robertson-Walker metrics. The geodesics of the spacetime are expressed as integral expressions which are used to examine the formation of multiple images and the observed shapes of non-point sources. We develop the lens mapping from the spacetime perspective, and use the Jacobian of the mapping to explain the observed image shapes. Approaching the geodesic equations as ordinary differential equations, we demonstrate the development of wave front singularities and time delays between light ray signals. This work demonstrates that the widely used thin lens approximation can be replaced with more robust techniques aligned with general relativity.

\end{abstract}

\maketitle


\section{Introduction} \label{intro:sec}

In Einstein's Theory of General Relativity, the geodesic equations govern the path of light rays through the spacetime, and the geodesic deviation equations indicate how nearby geodesics diverge or converge along the geodesic. From a theoretical standpoint, these equations are the basis for the observational science of gravitational lensing, which has developed as an important probe of cosmology. 

Historically, gravitational lensing played an important, well-documented role in convincing the scientific community that general relativity was the correct theory of gravity, starting with the observations led by Eddington in 1919 \cite{eddington}. Important figures in the history of astrophysics and relativity continued to think about possible observational consequences of gravitational lensing, including  Franz Zwicky in 1937 \cite{zwicky}, Albert Einstein in 1939 \cite{einstein39}, and Sjur Refsdal in 1964 \cite{refsdal}. The development of improved optics, including the CCD camera and space-based telescopes in the 1990's, accelerated observational studies of gravitational lensing. 

Virtually all observational gravitational lensing studies utilize a thin lens approximation in their analysis of multiple images as described in the seminal book by Schneider, Ehlers and Falco in 1992 \cite{ehlers}. Much current observational work uses the thin lens approximation to analyze the distortion of the shape of many background galaxies\cite{schneider}. Both styles of study allow astrophysicists to determine parameters of the gravitational lens object.

In the late 1990's, several authors began to reconsider gravitational lensing from a perspective consistent with relativity, absent the thin lens approximation. In \citet{ehlersNewman}, Jurgen Ehlers and Ted Newman introduced a general picture of lensing from the spacetime, or nonperturbative, point of view. Shortly thereafter, \citet{virbhadra} and \citet{frittelli} explored lensing in the Schwarzschild black hole geometry. Later, \citet{bozza02} and separately \citet{sereno2003}, examined lensing in the rotating, Kerr, geometry. A Living Review article by \citet{perlick} summarizes much of this work and provides important context.

Over the past 20 years, many other authors have considered exact lensing using the spacetime perspective and methods, often with the general hope of determining observational signatures in the lensing created by  different spacetime structures such as strings, or other defects. The hope is that these lensing signatures might provide a test of different possible gravity models, including several recent papers \cite{Lan, Hsieh, Chen}.

All of these studies work in the framework of exact gravitational lensing. While potentially important, these applications of the spacetime perspective are not applicable to the common observational context of gravitational lensing by galaxies or galaxy clusters co-moving in the universe. The purpose of this paper is to consider aspects of the spacetime perspective of gravitational lensing by gravity lenses embedded in cosmological solutions to the Einstein Field Equations.

While we will follow the spacetime perspective, we are unable to fully follow the exact lensing formalism common to previous papers. Fundamentally, the starting point of the spacetime perspective is the metric, and in studies one thinks of as exact lensing, the metric should be an exact solution of the Einstein Field Equation. The best one can do for a spacetime perspective of lenses embedded in cosmological metrics is to start with a perturbed metric. In section \ref{perturbedFRW:sec}, we outline the metric and perturbation we consider for this examination. We will refer to these metrics as cosmological lensing metrics. In this initial study of the subject, we consider only spherically symmetric perturbations to a flat, $\Lambda$-dominated Friedman-Robertson-Walker metric as this cosmological model is most preferred by astrophysical data.

Our current work expands on the work by \citet{futamase} who also studied lensing in similar spacetime metrics. Our approach will differ from Futamase by fully following the spacetime perspective approach, while Futamase's approach is to work from the metric to obtain a bending angle and then work within the traditional thin lens mapping approach.

This paper is organized as follows. First, in section \ref{perspective:sec}, we outline the spacetime perspective of gravitational lensing in the cosmological context, as compared with the thin lens approximation. Before considering perturbed cosmological metrics, we consider lensing by the cosmology itself, which leads to distortion of bundles of rays, in section \ref{dyer:sec}. After introducing our metric in section \ref{perturbedFRW:sec}, we consider the geodesic equations in section \ref{geodesics:sec} in a closed integral format, and we discuss light ray paths in section \ref{images:sec}. Section \ref{shapes:sec} shows the distortion of images, and section \ref{mapping:sec} develops the lens mapping from the spacetime perspective, allowing one to explain the image distortion based on the Jacobian of the lens mapping. In section \ref{time:sec}, we reconsider the geodesic equations as ordinary differential equations evolving in time, and use this approach to the geodesics to consider wave fronts in Section \ref{wavefronts:sec} and time delays in Section \ref{timedelays:sec} uses. 


\section{Thin Lens and Spacetime Perspectives of Lensing} \label{perspective:sec}

As a general rule, observational lensing studies make use of the thin lens approximation, where it is assumed that the path of the light ray between the source and observer are two connected straight lines that meet at the lens plane. At the lens plane, the light ray changes direction instantaneously in the amount of a ``bending angle'' which one computes from general relativity. The thin lens approximation is frequently formulated as a mapping between the lens plane (positions in a 2-d plane perpendicular to the line of sight from the observer to the center of the lens) and the source plane (positions in a 2-d plane perpendicular to the line of sight connecting the observer and lens at an angular diameter distance $D_d$ away from the observer). The thin lens approximation is justified in the sense that the distance scale at which the gravity is substantial is small compared to the distance scales between the lens and observer and the lens and light source. 

In the typical thin lens approach, including \citet{ehlers}, people use the approximation that the gravitational potential depends on the proper distance at the time the light ray passes the lens. Thus, the entire input of the cosmological background in the thin lens perspective is hidden in two places: in the use of angular diameter distances which depend on the cosmological model and the source and lens redshifts, and in the assumption that the gravity bending depends on the proper distances in the lens plane.

In the spacetime perspective, the key structure is the past light cone of the observer at the moment of observation. The apex of the light cone is spanned by a two parameter family of initial directions which we can think of as the sky. The equations of gravitational lensing are represented as a mapping from the two dimensional sky of the observer to positions of sources. 

From the spacetime perspective, the geodesic equations connect the observer to the light sources on the observer's past light cone, playing the role of the straight lines connected by a deflection angle in the thin lens approximation. An extended source of light can be modeled as a bundle of light,  where the central line connecting the observer to the center of the source is a central geodesic, and points along the edge of the bundle are connected to the central geodesic by geodesic deviation vectors. Alternatively, the edge of an extended source sends light rays that also meet at the observer's light cone's apex. 

As a result of the curvature of the spacetime, there may be more than one geodesic that connects the observer and source, which is often referred to as strong lensing. Further, the curvature of the spacetime might cause distortion to the bundle of light rays either through shear or convergence. Through the geodesic deviation equations, the shearing and convergence (or divergence) of the light ray bundle leads to image shape distortion or magnification, which are often referred to as aspects of weak lensing.  


\section{Lensing by non-perturbed cosmologies}\label{dyer:sec}

It is well known that cosmological distances are influenced by the geometry of the expanding universe. Particularly important in gravitational lensing is the angular diameter distance, which is used in the thin lens approximation to create the lens equation and lens mapping. Within the gravitational lensing community, the angular diameter distance's behavior as a function of redshift is determined by the Dyer-Roeder equation \cite{ehlers}.

While it is not usually thought of as gravitational lensing, from the spacetime perspective, the variation of the angular diameter distance, as well as its cousin the luminosity distance, with redshift or look-back time are a gravitational lensing effect. This is because the angular diameter distance is fundamentally an expression of geodesic deviation. In fact, one can derive the Dyer-Roeder equation directly and easily from the geodesic deviation equation expressed in the Newman-Penrose formalism \cite{aly}.

Hence, from the spacetime perspective, we consider the variation of angular diameter distances with redshift or look-back time as an expression of gravitational lensing by the cosmological spacetime. In this case, the whole of the curvature of the spacetime is the lens. When a separate lens is introduced in the next sections of this paper, the impact of the cosmology on the light rays remains, and this baseline gravitational lensing by the underlying cosmology is added to by the lensing of the perturbation.


\section{Lensing Model} \label{perturbedFRW:sec}

In this section, we set up the basic lensing scenario under consideration. We assume a spherically symmetric mass density that represents a typical galaxy cluster at moderate redshift.

\subsection{Cosmological Metric} 
For the remainder of this paper, we will assume that the physics for gravitational lensing in the context of an expanding universe is given by a metric that represents a weak gravitational field perturbation of a flat Friedman-Robertson-Walker cosmological metric. When we consider wave fronts, it is easier to describe the spatial part of the metric using Cartesian coordinates, but for our initial discussion, we begin with spherical coordinates and specify the line element of our cosmological metric as

\begin{equation} ds^2 = \left(1 + 2 \varphi \right) dt^2 - a^2(t) \left( 1 - 2 \varphi \right) \left( dr^2 + r^2 d\theta^2 + r^2 \sin^2\theta d\phi^2 \right) . \label{perturbMetric:eqn} \end{equation}

\noindent Here, the gravitational potential of the lens is represented by $\varphi$, and the cosmic expansion is given by $a(t)$. We have suppressed factors of the speed of light $c$, and assume the $\varphi$ is small. The Einstein Field Equation for this metric, working to first order in $\varphi$ and assuming that $\dot a / a \times \varphi \ll \varphi$, results in a Poisson equation for $\varphi$ and the usual Friedman equations for $a(t)$. Because the metric itself is approximate, and accurate only to first order in $\varphi$, all equations in this paper are derived excluding terms of order $O(\varphi^2)$ or higher unless specified.

Taking the overall universe as a flat with matter and a positive cosmological constant, the Friedmann equation results in the scale factor taking the form

\begin{equation}
    a(t) = \left( \frac{\Omega_{m,0}}{\Omega_{\Lambda,0}} \right)^{1/3} \left\{ \sinh \left( \frac{3 H_0 \sqrt{\Omega_{\Lambda,0}} t}{2} \right) \right\}^{2/3}. \label{a(t):eqn}
\end{equation}

\noindent We will take the energy densities to have the numerical values $\Omega_{m,0} = 0.31$ and $\Omega_{\Lambda,0} = 0.69$ , with  the Hubble constant $H_0 = 68$~km s$^{-1}$~Mpc$^{-1}$ \cite{Ryden}. Assuming the current age of the universe to be $t_o = 13.7$ billion years, $a(t_o) = 1.00122$.

\subsection{Lens Matter Density}

Throughout this paper, we will assume that the gravitational potential is spherically symmetric and centered at the origin. Following standard practice, we will assume that the gravitational potential depends on the proper distance at the time the light ray passes the lens \cite{ehlers}. If the lens redshift is $z_l$, then $\varphi(r_p)$ with 

\begin{equation} r_p = \frac{a(t_o)}{a(t_l)} r = \frac{a(t_o)} {1+z_l} r. \end{equation}

As a standard lens, we consider a truncated isothermal sphere model with a core radius, whose three dimensional mass density is given by 

\begin{equation} \rho = \frac{\sigma_v^2}{2\pi G}\frac{1}{r_c^2 + r_p^2} \frac{r_t^2}{r_t^2 + r_p^2}. \label{SISCoreDensity:eqn}
\end{equation}

\noindent With both the core radius, $r_c$, and truncation radius, $r_t$, set to zero, this model is the commonly used Singular Isothermal Sphere model. The SIS model predicts flat rotation curves, but has a singular density at the origin and an infinite total mass. The introduction of the core radius $r_c$ maintains a finite density at the origin, and the truncation scheme used here, the fraction $r_t^2 / (r_t^2 + r_p^2)$, ensures that the mass density falls off quickly after some large truncation radius, leading to a finite total mass. This density is modeled off a similar truncated Navarro, Frenk, and White (NFW) model introduced by \citet{baltz} and studied in the context of gravitational lensing by \citet{campbell}.

For our model in this paper, we assume the velocity dispersion $\sigma_v$ to have the value of 1500 km s$^{-1}$, core radius $r_c = 250$~kpc and truncation radius $r_t = 1.5$~Mpc. With these values, the total mass of the modeled galaxy cluster within one Mpc is $8.7 \times 10^{14}$ solar masses, and the total mass of the cluster is about $2.4 \times 10^{15}$ solar masses. This is consistent with a large gravitationally bound cluster of galaxies.

\subsection{Initial Conditions and Angular Diameter Distances}

We will assume that the lens is at a redshift of $z_l = 0.45$, with typical background sources averaging a redshift of $z_s = 0.8$. These values, combined with the lens parameters above, model the cluster RX J1347-1145, which has two visible strong lensing arcs at about $35$ arc seconds from  a very distant source \cite{rxj2} and a measurable weak gravitational lensing signal \cite{rxj}.

To place the observer and source, we work with the flat cosmology metric and consider radial light rays. From the definition of the redshift, $z_r$

\[ \frac{1}{1+z_r} = \frac{a(t_o)}{a(t_e)}\]

\noindent and the scale factor, Eq.~\ref{a(t):eqn}, for both the lens and source redshifts, we solve for the times of emission $t_e$ for light sources at redshifts $z_l = 0.45$ and $z_s = 0.80$. Then, using $ds^2=0$ for light rays, we numerically integrate

\begin{equation} \int_{t_o}^{t_e} \, \frac{1}{a(t)} dt = \int_0^{r_o}\, dr \end{equation}

\noindent to find the radial coordinates of the observer. A similar integral is done for the radial location of the sources. We then place the center of the lens at the origin, and, using the integrals, obtain $r_o = 5.7404$~billion years for the initial position, representing the apex of the past light cone of the observer and $r_f = 3.533$ billion years as the final radial position of sources. 

Angular diameter distances are needed in when considering the Jacobian of the lens mapping. We follow the techniques presented in \citet{hogg}, where the angular diameter distance to a source at redshift $z_s$ is given by 

\[ D_s = \frac{1}{1+z_s} \times D_c(z_s). \]

\noindent For the co-moving distance in the flat cosmological model considered in this paper,

\begin{equation} D_c(z_r) = \frac{c}{H_o}\frac{1}{1+z_r}\int_0^{z_r} \, \frac{d z'_r}{\sqrt{\Omega_\Lambda + (1+z'_r)^3 \Omega_m}}. \end{equation}

\noindent The angular diameter distance between the source and lens would be given by 

\begin{equation}  D_{ds} = \frac{c}{H_o} \frac{1}{1+z_s} \left(D_c(z_s) - D_c(z_l) \right).  \end{equation}


\section{Equations for Light Ray Geodesics} \label{geodesics:sec}

When considering the paths of light rays, it is convenient to transform the cosmological metric through a conformal time transformation. We begin with our cosmological lensing metric, Eq.~\ref{perturbMetric:eqn}, and introduce a new time variable defined by

\begin{equation}
    d \tilde t = \frac{dt}{a(t)}. \label{conformaltime:eqn} 
\end{equation}

\noindent By integrating both sides of this equation, we can solve for $\tilde t$ as a function of $t$, and then with this conformal time, the metric becomes

\begin{equation}
    ds^2 = a^2(\tilde t) \left\{  \left(1 + 2 \varphi \right) d{\tilde t}^2 - \left( 1 - 2 \varphi \right) \left( dr^2 + r^2 d\theta^2 + r^2 \sin^2\theta d\phi^2 \right) \right\}. 
\end{equation}

\noindent We are then able to work in a new metric, conformally related to our cosmological lensing metric, given by

\begin{equation}
    d{\tilde s} ^2 = \frac{1}{a^2} ds^2 = \left(1 + 2 \varphi \right) d{\tilde t}^2 - \left( 1 - 2 \varphi \right) \left( dr^2 + r^2 d\theta^2 + r^2 \sin^2\theta d\phi^2 \right) \label{conformalMetric:eqn}.
\end{equation}

The conformal metric, Eq.~\ref{conformalMetric:eqn}, presents several advantages for examining the spacetime perspective of cosmological lensing. First,  the conformal metric is time independent when one assumes that the gravitational potential, $\varphi$, depends only on the physical distances on the spatial hypersurface at the time the light ray passes our gravitational lens. Second, the light rays of the conformal metric follow the same spatial paths through the $(r, \theta, \phi)$ or $(x,\,y,\,z)$ coordinates as they would in the cosmological lensing metric. Finally, the angles between spatial vectors in both metrics are identical, so the directions of observation would be the same.

Working with the conformal cosmological lensing metric, we introduce a Lagrangian defined by 

\begin{equation}
    {\mathcal{L}} = \frac{1}{2} \left\{ \left(1 + 2 \varphi \right) {\dot{\tilde t}}^2 - \left( 1 - 2 \varphi \right) \left( \dot r^2 + r^2 \, \dot \theta^2 + r^2 \sin^2\theta \, \dot \phi^2 \right) \right\}, \label{conformalL:eqn}
    \end{equation}

\noindent where the dot represents a derivative with respect to the parameter $\tilde s$. To begin our studies, we make use of the symmetry of the lens around the axis connecting the center of the lens and the observer to assume the light rays travel in a plane such that $\phi$ is fixed at $\phi = 0$. Translating to Cartesian coordinates, our light rays are confined to the $x-z$ plane. With this assumption, the $\tilde t$ and $\theta$ variables are absent from the Lagrangian, and there are first integrals of motion for those coordinates.  Working to first order in the gravitational potential, we have

\begin{eqnarray}
    \dot{\tilde t} &=&  -  \left(1  -2 \varphi \right), \nonumber \\
    \dot{\theta} &=&   \frac{b}{r^2} \left(1 + 2 \varphi \right). \label{firstIntegrals:eqn}
\end{eqnarray}

\noindent The choice of signs on the right hand side of each equation simplifies the equations and ensures that as the parameter $\tilde s$ increases, the time coordinate runs backwards.

To find a third equation for the $r$ coordinate, it is convenient to use the fact that we are solving for light rays, in which case

\begin{equation}
    {\mathcal{L}} = \frac{1}{2} g_{ab} \dot x^a \dot x^b = 0.
\end{equation}

\noindent We therefore set the conformal cosmological lensing Lagrangian, Eq.~\ref{conformalL:eqn}, to zero, and solve for $\dot r$. We simplify results using the two first integrals in Eqs.~\ref{firstIntegrals:eqn} and work to first order in the potential $\varphi$. The result is

\begin{equation}
    \dot r = \pm \sqrt{1 - \frac{b^2}{r^2} \left( 1 + 4 \varphi \right)} \label{rdot:eqn}.
\end{equation}

\noindent In Eq.~\ref{rdot:eqn}, the minus sign is used for the portion of the light ray trajectory where the light ray approaches the lens located at the origin. The light ray will reach a point of closest approach, $r_{m}$, given by the largest root of the term under the square root. Because as the light ray approaches the point of closest approach the entire term inside the square root goes to zero, it is not possible to simplify the expression by assuming the $\varphi$ is small. After reaching the point of closest approach, the $r$ coordinate will increase and the plus sign is appropriate. 

Closed form expressions for the path of the light ray are given by combining $\dot \theta$ and $\dot r$ expressions from Eqs.~\ref{firstIntegrals:eqn} and \ref{rdot:eqn}. Defining

\begin{equation}
    \frac{d\theta}{dr} = \frac{d\theta}{d\tilde s} \frac{d\tilde s}{dr},
\end{equation} 

 \noindent we can form an integral expression for the value of $\theta$ as a function of the $r$ coordinate. We will assume that the observer sits at $\theta = \pi$ and $r = r_o$. This represents the observer sitting on the $-z$ coordinate axis. For points along the light ray between the observer at $r_o$ and the point of closest approach, we have

 \begin{equation}
     \theta(r) = \pi - \int_{r_o}^{r}  \frac{b \left(1 + 2 \varphi \right) }{r^2 \sqrt{1 - \frac{b^2}{r^2} \left( 1 + 4 \varphi \right)} } dr. \label{thetaInt_incoming:eqn}
 \end{equation}

\noindent For rays that have passed the point of closest approach, $r_m$, we would have

 \begin{eqnarray}
     \theta(r)  &= & \pi - \int_{r_o}^{r_m}  \frac{b \left(1 + 2 \varphi \right) }{r^2 \sqrt{1 - \frac{b^2}{r^2} \left( 1 + 4 \varphi \right)} } dr \nonumber \\  &~&~ + \int_{r_m}^{r}  \frac{b \left(1 + 2 \varphi \right) }{r^2 \sqrt{1 - \frac{b^2}{r^2} \left( 1 + 4 \varphi \right)} } dr . \label{thetaInt_full:eqn}
 \end{eqnarray}

\subsection{Relation between $b$ and observation angle}

The parameter $b$ was a constant of integration in the $\theta$ Euler-Lagrange equation, Eq.~\ref{firstIntegrals:eqn}. As we see in Eq.~\ref{thetaInt_full:eqn}, the $b$ parameter determines the final $\theta$ value of the source location. We can relate $b$ to an angle between the observer and the optical axis, or the $\hat z$ axis, by considering the inner product of the spatial part of the tangent vector to the light ray, $\ell^i$ and a spatial vector that points radially towards the origin $o^j$. We have

\begin{equation} 
    \cos\psi = \frac{g_{ij} \ell^i o^j}{\sqrt{|g_{ij} \ell^i \ell^j |} \sqrt{|g_{ij}o^i o^j|}}, \label{psiDef:eqn}
\end{equation}

\noindent for $\ell^i = (\dot r, \dot \theta, 0)$ and $o^i = (1, 0, 0)$ with the spatial part of the metric given by Eq.~\ref{conformalMetric:eqn}. We evaluate the $r$ coordinate at $r_o$, and then using $\dot \theta$ from Eq.~\ref{firstIntegrals:eqn} and $\dot r$ from Eq.~\ref{rdot:eqn}, work to first order for small $\varphi$ and small $\psi$, to obtain a relation between $b$ and the angle the observer sees for the light ray $\psi$:

\begin{equation}
    \psi = \frac{b}{r_o}. \label{obsAngle:eqn}
\end{equation}

\section{Light Ray Paths and Multiple Images} \label{images:sec}

Throughout this paper, we assume the center of the lens is at the origin, and the observer is located at a point $r_o$ and $\theta = \pi$ so that the observer is positioned on the $\hat z$ axis, which we refer to as the optical axis. We take the lens at the redshift $z_l$ as described above. We will consider different sources of light that are assumed to be at $z_s  = 0.8$, which is converted to a radial position, $r_f$. 

In this scenario, the $b$ parameter, initially introduced in Eq.~\ref{firstIntegrals:eqn}, determines the direction of the past-directed light ray at the observer so that $b$ parametrizes the sky of the observer. Setting $r=r_f$ defines the source plane, or a two-dimensional spatial surface where sources (at a given redshift) are located. We will use Eq.~\ref{thetaInt_full:eqn} with the final $r$ limit set to $r=r_f$ to determine the position of a source, and convert those positions to Cartesian coordinates for plotting using the usual trigonometric relations. We can re-introduce the third spatial dimension using a free rotation in the $\phi$ coordinate. For this reason, we think of a radial coordinate distance $s = r_f \sin\theta \approx r_f \theta $ as the distance of the source from the optical axis.

\begin{figure}
\begin{center}
\scalebox{1.0}{\includegraphics{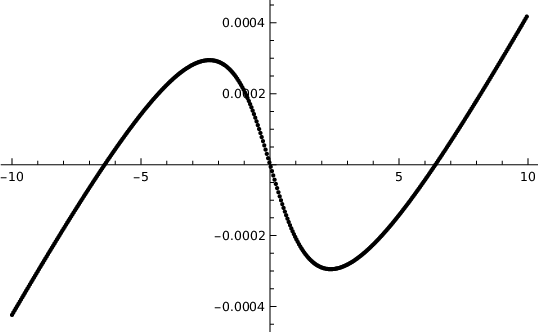}} \caption{\label{xversusb:fig}
The final coordinate distance from the optical axis as a function of the initial parameter $b$. Plotted are $s=r_f\sin\theta$ in billions of years versus $b/r_c$ where $r_c$ is the core radius in the mass model. For large positive and negative values of $b$, there is only one light ray that connects the source and observer. As a horizontal line becomes tangent to the local extrema, the source, at $s$ passes through a caustic and multiple images appear.} \end{center}\end{figure}

The light rays we are interested in must pass close to the gravitational lens, making the final source position's $s$ very small compared with the $r_f$ values. Figure \ref{xversusb:fig} shows the final $s$ values as a function of the initial $b$ values. In this figure, the vertical axis is plotted in units of billions of light years, and the horizontal axis represents the value of $b/r_c$, where $r_c$ is the core radius in the mass model. For large positive or negative $s$ values, we see that the mapping from $b$ to $s$ (or from the sky of the observer to the source plane) is one-to-one. 

However, we see that there is a region for which three different $b$ values correspond to the same source location. This occurs when a horizontal line becomes tangent to the local minimum (or maximum) of Fig.~\ref{xversusb:fig}. Because of the spherical symmetry of the lens, there is axial symmetry around the optical axis connecting the observer and the origin, which is the spatial center of the lens's mass distribution. The coordinate radius at the point of tangency in Fig.~\ref{xversusb:fig} becomes a circle in the source plane which we will interpret as the caustic curve of the lens mapping.

Figure~\ref{4rays:fig} shows four light rays at different values of $b$. One light ray, at $b=9.81\,r_c$, or $9.81$ times the core radius value, is in the region of Fig.~\ref{xversusb:fig} where only one light ray connects the observer to the source location.  The other three rays, at $b=-4.33\,r_c$, $b=-0.95\,r_c$, and $b = 8.16\,r_c$ correspond to rays that all connect the observer to the same source location at a coordinate radius of $s = 0.0002$ billion years.

\begin{figure}
\begin{center}
\scalebox{0.7}{\includegraphics{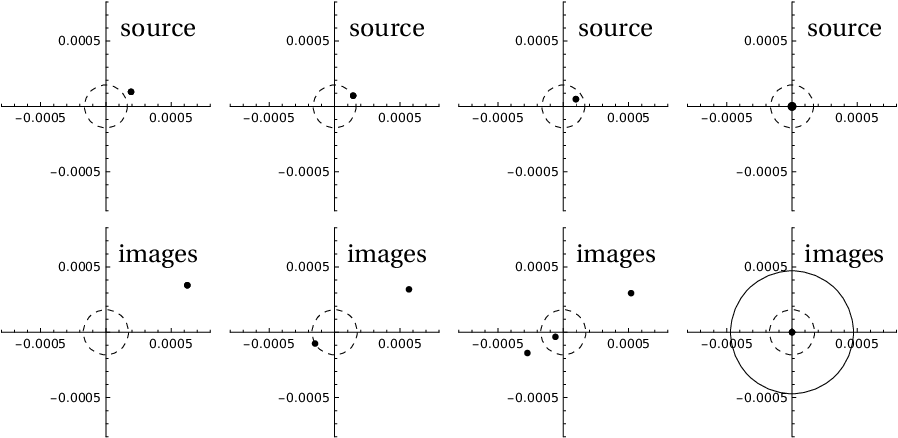}} \caption{\label{gridImages:fig}
Four source locations (top panel) and observed images (bottom panel) of a point source at locations along a line with $\phi = \pi/6$. In the source plane, the circular caustic is shown as a dashed curve. The dashed circle in the image plane represents the critical curve, or image locations that map to the caustic. In the second panel from the left, we see that when a point source is on caustic, a second image appears at the critical curve. An Einstein Ring appears if the source is at the origin in the source plane. Source and image locations in this plot are the physical distances from the optical axis in billions of years. } \end{center}\end{figure}

The possible image locations as a function of the locations of point sources are demonstrated in Fig.~\ref{gridImages:fig}, where upper panels represent source locations and lower panels show the corresponding image locations. The dashed circle in the upper panels is the caustic of the lens mapping in the source plane, and the dashed circle in the lower panels is the critical curve of image points that map into source positions on the caustic. We see that as the source crosses the caustic, a pair of images appears. As the source moves closer to the origin (on the optical axis), one of these images moves closer to the origin and the other moves out to the Einstein Ring radius. When the source is on the optical axis,  or at the origin of the source plane, symmetry creates a full Einstein Ring as shown in the bottom right panel.

\section{Shapes of Images} \label{shapes:sec}

We can use the integration of individual light rays, from Eq.~\ref{thetaInt_full:eqn}, to determine the observed shapes of extended sources. Our first step is to create a set of points that represent the boundary of the extended source object. We start by picking a coordinate distance for which there would be three images in Fig.~\ref{xversusb:fig}. For this example, we use $s = 0.0002$ billion light years, and rotate the $x$ axis through a $\phi$ angle of $\phi_c = \pi/6$. This allows us to set the center of our extended object at

\begin{eqnarray}
    x_c &=& s \cos \phi_c, \nonumber \\
    y_c &=& s \sin \phi_c. \nonumber
\end{eqnarray}

\noindent The extended object, with a circular profile would then be bounded by edge points

\begin{eqnarray}
    x_E &=& x_c + r_E \cos\alpha_E, \nonumber \\
    y_E &=& y_c + r_E \sin\alpha_E, \nonumber
\end{eqnarray}

\noindent where here the $E$ subscript refers to a quantity associated with one of the edge points. To make a figure that visualizes the effect of lensing on an extended object, we used $18$ points around the central object, spacing $\alpha_E$ evenly between $0$ and $2\pi$. We took $r_E$, the coordinate radial size of our extended object, to be $r_E = s / 10$, which is far too large to represent a physical galaxy, but is needed to see the effect on a plot. 

From the $(x_E, y_E)$ pairs, we can compute both a coordinate distance from the $(x,y)$ origin, $q_E$, and a rotation angle, $\phi_E$, relative to the origin. By setting $\theta_E = q_E / r_f$ with $r_f$ equal to the radial coordinate distance to source plane, we can then treat Eq.~\ref{thetaInt_full:eqn} in reverse, where the final $\theta$ is set, and we solve for the three $b$ values, $b_I$, resulting in that $\theta = \theta_E$. 

The image of the extended object is obtained by using Eq.~\ref{thetaInt_full:eqn} for each of the three $b_I$ with $\varphi = 0$ for no lensing, integrating out to $r_f$ to obtain a $\theta_I$. Then with $\phi_E$ from above we determine the edge of the image using the small angle approximation for $\theta_I$ as

\begin{eqnarray}
     x_I &=& (r_f \theta_I) \cos\phi_E, \nonumber \\
     y_I &=& (r_f \theta_I )\sin\phi_E. \nonumber
\end{eqnarray}

\begin{figure}
\begin{center}
\scalebox{1.0}{\includegraphics{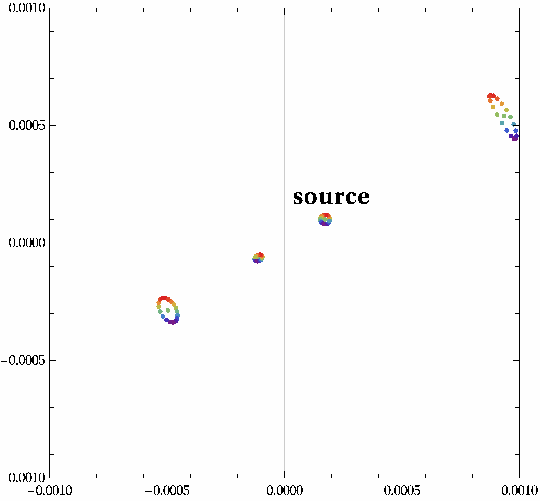}} \caption{\label{3images:fig}
The observed images of a circular extended source at $s = 0.0002$ billion years and $\phi = \pi/6$. All images are projected to the source plane so that the relative sizes of the images and source indicate whether the image is larger or smaller than the source. The outer images are larger than the source and elliptically stretched tangent to a circle centered on the optical axis, which runs through the origin in this figure. The more central image is stretched along the radial line and is visibly smaller than the source. Colors of points on the extended source and on each image show how the points along the edge of the source rotated. Axes denote coordinate distances in billions of light years.} \end{center}\end{figure}

Figure \ref{3images:fig} shows the result of this process for the initial $s=0.0002$ billion years and $\phi_c = \pi/6$ with $r_E = s/10$. We plot $19$ points, one at the center and $18$ edge points. Because we integrated the $\theta_E$ all the way to the source plane, the scale of the three images and source are the same, and visual changes in size correspond to actual observable differences. Shown are the initial circular source and three images. The outer two images are stretched into ellipses whose long axis is tangent to circles centered along the optical axis, and they are larger than the original source in area. The lensing caused the inner image, appearing close to the origin in the third quadrant, to be stretched into an elliptical shape whose long axis points towards the origin. This image is smaller than the original source.

\begin{figure}
\begin{center}
\scalebox{1.0}{\includegraphics{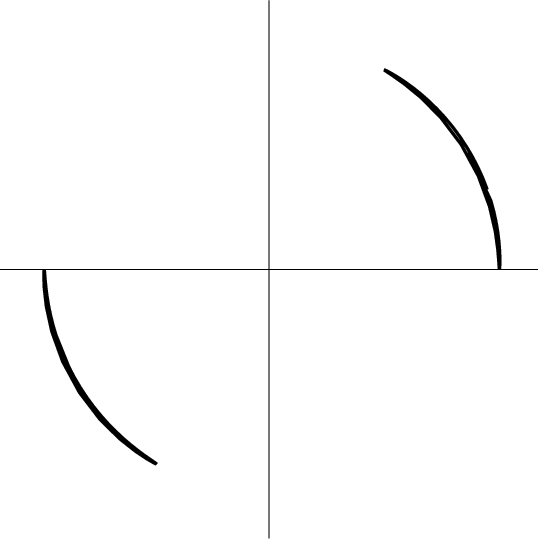}} \caption{\label{partialRing:fig}
The outer two observed images of a small point source near the optical axis connecting the observer and center of the lens. The coordinate radius of the center of the light source was set to $s= 0.0001$ billion years with $\phi = \pi/6$. The images are highly stretched along the circle centered on the optical axis and begin to form a complete Einstein Ring.} \end{center}\end{figure}

The sudden appearance of the Einstein Ring in Fig.~\ref{gridImages:fig} for the point source is better understood when looking at small extended sources of non-zero area. Figure~\ref{partialRing:fig} shows the image of a small circular source very close to the origin in the source plane. We see two extended arcs that stretch substantially and will connect into a full Einstein Ring as the source moves ever closer to the optical axis.

\section{Lens Mapping, Jacobian Matrices, and Image Distortion}\label{mapping:sec}

In the traditional thin lens approach, one considers the lensing action as a mapping from a two-dimensional lens plane into a two dimensional source plane. When there are multiple images of the same source, the mapping fails to be one-to-one. The locations in lens plane where the determinant of the Jacobian of the lens mapping is zero are critical curves, and the projection of the critical curves into the source plane are caustic curves which separate regions where there can be different numbers of images of the same source \cite{ehlers}. 

The observed shape of a small, extended source is determined from the Eigenvectors and Eigenvalues of the Jacobian matrix. A small circular source is stretched or compressed along the Eigenvectors by an amount inversely proportional to the absolute value of the Eigenvalue.

Some authors have formulated the spacetime perspective of the lens mapping as a mapping from a two-dimensional space of parameters that span the initial directions of light rays of the observer's past light cone (the ``sky of the observer'') into a two-dimensional space corresponding to (possibly a portion of) the source plane \cite{ehlersNewman, virbhadra, frittelli}. Following this approach, since our observer is on the optical ($\hat z$) axis about which there is axial symmetry, we can think of the final source positions at $r_f$ as the pair $(\theta, \phi)$ with $\theta$ given by Eq.~\ref{thetaInt_full:eqn} and $\phi$ being a free parameter. Since $\theta$ is determined by the parameter $b$ and $\phi$ is a free rotation about the $\hat z$ axis, the lens mapping in this version of the spacetime perspective of lensing is a mapping from the space spanned by $(b, \phi)$ into the space of $(\theta, \phi)$. The Jacobian matrix would then be the matrix of partial derivatives of $\theta$ and $\phi$ with respect to $b$ and $\phi$.

However, as we will see, explaining the shape of extended images shown in Section~\ref{images:sec}, requires us to consider the spacetime lens mapping in terms of mappings of physical distances from the optical axis. To do this, we note that the $\theta$ coordinate in our lens mapping, Eq.~\ref{thetaInt_full:eqn}, is invariant under the conformal transformation we made in the original metric. Another way of saying this is that the path through the $(r,\theta,\phi)$ space in the conformal metric is the same as the path of a light ray making the same angle with the optical axis in the original cosmological metric, Eq.~\ref{perturbMetric:eqn}. Physical distances in the source plane are then determined by the angular diameter distances influenced by the cosmological metric's expansion in time.

A ray parametrized by a given $b$ value would then intersect the source plane at $r_f$ at a angular position $\theta$, from Eq.~\ref{thetaInt_full:eqn}, which we might rotate through an angle $\phi$. The physical position of the source would be

\begin{eqnarray}
    x_s &=& D_{ds} \sin\theta \cos\phi, \nonumber \\
    y_s &=& D_{ds} \sin\theta \sin\phi, \label{source:eqn}
\end{eqnarray}

\noindent using the angular diameter distance between the lens (or deflector) and source plane $D_{ds}$. The position of the image depends on the angles $\psi$ and $\phi$ visible at the observer and the angular diameter distance to the source plane, $D_s$:

\begin{eqnarray}
    x_I &=& D_s \sin\psi \cos\phi\, \nonumber \\ 
    y_I &=& D_s \sin\psi \sin\phi. \label{image:eqn}
\end{eqnarray}

In the spacetime perspective considered in this paper, it is most appropriate to take the lens mapping as a mapping from $(x_I, y_I)$ to $(x_S, y_S)$. The desired Jacobian matrix is

\begin{equation}
    J = \left( \begin{array}{cc} \frac{\partial x_S}{\partial x_I} & \frac{\partial x_s}{\partial y_I} \\ \frac{\partial y_S}{\partial x_I} & \frac{ \partial y_S}{\partial y_I} \end{array} \right). \label{Jacob:eqn}
\end{equation}

\noindent Because the variables we control are $b$, which is related to $\psi$, and the free rotation $\phi$, we would like to express the Jacobian matrix in terms of derivatives with respect to $b$ and $\phi$. Noting that $\sin\psi = \sqrt{x_I^2 + y_I^2}/D_s$ and $\tan\phi = y_I/x_I$, we have 

\[ \frac{\partial}{\partial x_I} = \frac{x_I}{D_s \sqrt{x_I^2 + y_I^2}} \frac{\partial}{\partial \psi} - \frac{y_I}{x_I^2 + y_I^2} \frac{\partial}{\partial \phi}, \]

\noindent with a similar expression for $\partial/\partial y_I$. The Jacobian matrix can then be expanded in terms of these derivatives in terms of the angles $\psi$ and $\phi$. Simplifications occur in the case of axial symmetry. After computing the derivatives, one can use axial symmetry and impose $\phi = 0$.  A small $\theta$ approximation results in a Jacobian matrix

\[ J = \left( \begin{array}{cc} \frac{D_{ds}}{D_s}\frac{\partial \theta}{\partial \psi} & 0 \\ 0 & \frac{D_{ds}\theta}{D_s \psi} \end{array} \right).
\]

\noindent While this Jacobian matrix applies to points sources and images along the $\hat x$ axis (for $\phi = 0$), the axial symmetry allows us to take results from this Jacobian matrix and apply them to other positions at the same distance from the central axis through rotations. Because $\theta$ from Eq.~\ref{thetaInt_full:eqn} is directly a function of $b$, we compute the $\psi$ derivative via the chain rule and use Eq.\ref{psiDef:eqn} to simplify the resulting Jacobian matrix:

\begin{equation} J = \left( \begin{array}{cc} \frac{r_o\, D_{ds}}{D_S} \frac{\partial \theta}{\partial b} & 0 \\ 0 & \frac{r_o D_{ds}\theta}{D_s b} \end{array} \right). \label{Jacob_axis:eqn}
\end{equation}

\noindent Because the Jacobian matrix is diagonal at $\phi=0$, the two Eigenvectors will point in the horizontal, or $\hat x$, and vertical, or $\hat y$, directions.

Table \ref{shapes:table} uses the Eigenvectors and Eigenvalues of the Jacobian matrix, Eq.~\ref{Jacob_axis:eqn}, to explain the shape of the observed images of the circular source. The derivative in the Jacobian matrix was computed numerically. Three sources, located a coordinate distance $s = r_f \theta$ from the optical axis, are considered. The first source, at $s = 0.0004$ billion years, has only one image. For that image, the Eigenvalue corresponding to the horizontal Eigenvector, the one parallel to the $\hat x$ axis, was $0.94$, and the Eigenvalue for the vertical Eigenvector, parallel to the $\hat y$ axis, was $0.31$. As a result, the ratio of the absolute values of the inverse of the Eigenvalues is $3.04$; this is listed as the stretch in the table. This means that the image of a circular source will be a vertically oriented ellipse whose vertical axis is about three times its horizontal axis in length. The product of the absolute values of the inverse of the Eigenvalues was $3.45$, which is listed as the area in the table. This means that the area of the image will be $3.45$ times larger than the original source's area. Generally, as the source position becomes further away from the optical axis, the images become more round and more similar in area to the original source; both the stretch and area approach one.

For source locations where there are three images, how close the source is to the caustic location has implications for the shapes of the images. We saw, in Section~\ref{shapes:sec}, that a circular source at $s = 0.0002$ billion years had three images, where the two outer images were elliptical, and stretched along a circle centered on the optical axis, but that the central image was elliptical and stretched along the line connecting the source and optical axis. In Table~\ref{shapes:table}, we see that for this source location, the furthest right image will have a large vertical stretch ($4.85$) and a large area ($6.00$). The leftmost image is slightly smaller, with a smaller vertical stretch. The middle image is in fact smaller than the source (area $=0.54$) and has a larger stretch along the axis pointing towards the optical axis than perpendicular to it (stretch $=0.73<1$), consistent with Fig.~\ref{3images:fig}. As the source location approaches the optical axis, the middle image becomes more circular with stretch closer to one, but also smaller, with area approaching zero. 

\begin{table}[]
    \centering
    \begin{tabular}{c|c|c|c}
         $r_f \theta$ & Location & Stretch & Area \\
         \hline
         ~ & left & 5.33 & 9.69 \\ 
         0.0001 & middle & 0.94 & 0.33 \\
         ~ & right & 8.34 & 11.17 \\
         \hline
         ~ & left & 1.73 & 4.73 \\
         0.0002 & middle & 0.73 & 0.54 \\
         ~ & right & 4.85 & 6.00 \\
         \hline 
         0.0004 & right & 3.04 & 3.45
    \end{tabular}
    \caption{The vertical stretch and scaled area of images from gravitationally lensed sources. The coordinate distance $s = r_f \theta$ is the distance from the optical axis in the source plane in billions of years. The choice of three source positions shows three horizontal cuts of Figure~\ref{xversusb:fig}. An image with no distortion would have stretch and area values of 1. Stretch greater than one indicates the image is stretched perpendicular to the line connecting the optical axis and source. }
    \label{shapes:table}
\end{table}

\section{Wave Fronts and Time Delays in the Cosmological Lensing} 
\label{time:sec}

Our approach to cosmological lensing so far has followed a standard practice within the spacetime approach by making use of integral expressions to determine the paths of light rays. Because, as yet, we have been uninterested in the time along the light rays, we have made use of a conformal metric, Eq.~\ref{conformalMetric:eqn}, that allows for this integral approach. 

Two important aspects of gravitational lensing require knowledge of the time of flight for individual rays. First, in the absence of lensing, temporal cross-sections of the observer's light cone will be spheres. However, the effect of lensing causes the light cone to fold over itself and develop wave front singularities. Wave front singularities are connected to both the magnification and number of images observed, and point to important underlying mathematical features of gravitational lensing \cite{Petters}. Second, as the paths of light rays connecting the source and observer are not the same length, and also are impacted by the gravitational time delay inherent to general relativity, there can be differences in the time of reception between signals. For instance, if a supernova occurs in the source galaxy, it will be seen first in one image and later in the others. 

To consider wave fronts or time delays, we must now return to the original cosmological metric given by Eq.~\ref{perturbMetric:eqn}. We will continue to assume that the potential is dependent on the proper distance at the time the light passes near the lens, or $\varphi(r_p)$ with $r_p = a(t_o) r/(1+z_l)$. However, now the metric has explicit time dependence in the scale factor $a(t)$, and there is no longer a conserved quantity associated with the $t$ coordinate. As a result, closed form expressions for the $\theta$ coordinate as an integral of a function of $r$ are not possible. 

With this is mind, it makes sense to write the Lagrangian in Cartesian coordinates and pursue numerical solutions to the ordinary differential equations generated from the Euler-Lagrange equation. We take as our Lagrangian

\begin{equation} {\mathcal{L}} = \frac{1}{2} \left\{ (1 + 2\varphi) \dot t^2 - a^2(t) (1 - 2\varphi)( \dot x^2 + \dot y^2 + \dot z^2 )\right\}, \label{LagWithT:eqn} \end{equation}

\noindent where $\varphi$ is a function of 
\[ r_p = \frac{a(t_o)}{1+z_l}\sqrt{x^2+y^2+z^2}. \]

\noindent Consistent with previous sections, we will use the symmetry of the lens to place the motion of the light rays initially in the $\hat x - \hat z$ plane by holding $y = 0$ fixed. The Euler-Lagrange equations result in three, second-order,  ordinary differential equations:

\begin{eqnarray}
   \ddot t =  & = & -a a'(1-4\varphi)(\dot x^2 + \dot z^2) - 2 \dot t (\varphi_{,x} \dot x + \varphi_{,z} \dot z), \nonumber \\
   \ddot x & = & -2\frac{a'}{a} \dot t \dot x + \varphi_{,x} (\dot x^2 - \frac{\dot t^2}{a^2} - \dot z^2) + 2 \varphi_{,z} \dot x \dot z, \nonumber \\
   \ddot z & = & -2\frac{a'}{a} \dot t \dot z + \varphi_{,z} (\dot x^2 - \frac{\dot t^2}{a^2} - \dot x^2) + 2 \varphi_{,x} \dot x \dot z . \label{wavefrontODEs:eqn}
\end{eqnarray}

\noindent In Eq.~\ref{wavefrontODEs:eqn}, we denote $da/dt = a'$, and $\varphi_{,x}$ indicates a partial derivative with respect to the $x$ coordinate. The equations are accurate to first order in $\varphi$ and assume the product of $\varphi$ with its derivative to be second order. 

For initial conditions, we return to the Lagrangian, Eq.~\ref{LagWithT:eqn}, and note that the Lagrangian must remain zero for all time for light ray trajectories. At the observer, we take $x_o = 0, z_o = -r_o$ as  initial conditions  to be consistent with the integral approach used in previous sections. We assume $t_o$ to be the current age of the universe and $\dot t$ to be $-1$ for past direct light rays at the initial location. Because of the expansion of the universe, $\dot t$ will not remain $-1$ as we go backwards in time. Evaluating the Lagrangian at $(t_o, x_o, z_o)$ with $\dot t = -1$ and solving ${\mathcal{L}} = 0$ for $\dot z_o$ results in

\begin{equation}
    \dot z_o = \sqrt{\frac{1}{a^2(t_o)} (1+4\varphi_o) - \dot x_o^2 }. \label{initz:eqn}
\end{equation}

\noindent Thus, we consider $\dot x_o$ to be a free parameter controlling the direction of the light rays at the initial location, and take $\dot z_o$ from Eq.~\ref{initz:eqn} to ensure that the path is a null geodesic. The relationship between an initial angle, $\psi$, and $\dot x_o$ can be determined in the same way as before by taking the inner product of spatial part of the tangent to the light ray $\ell^a = (\dot t, \dot x, \dot y, \dot z)$ and the spatial vector $o^j$ pointing towards the origin as in Eq.~\ref{psiDef:eqn}.

One can confirm that the spatial, $(x, z)$, paths of the light rays integrated numerically from Eqs.~\ref{wavefrontODEs:eqn} are identical to the spatial paths determined from the integral approach, as for example Fig.~\ref{4rays:fig}. However, the advantage of the differential equation approach is that we can now examine how individual, or sets of, light rays are progressing through the spacetime as a function of the cosmological co-moving time variable $t$. 

\subsection{Wave fronts}
\label{wavefronts:sec}

\begin{figure}
\begin{center}
\scalebox{1.0}{\includegraphics{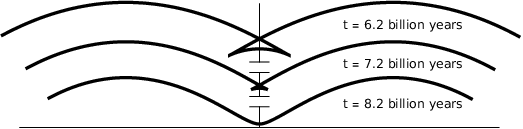}} \caption{\label{3waves:fig}
Portions of three wave fronts that have passed through the lens at successive times plotted along the $\hat x$ (horizontal) and $\hat z$ (vertical) axes. As the wave front moves into the past, it is traveling vertically upwards in this plot. The wave front folds over itself creating a cusp singularity. To show the evolution of the wave fronts on one graph, we have collapsed the vertical distance between the wave fronts, as shown by the parallel break lines. } \end{center}\end{figure}

Figure~\ref{3waves:fig} shows the portions of three wave fronts that have passed by the lens in the $\hat x-\hat z$ plane at $t = 8.2$ billion years, $t = 7.2$ billion years, and $t = 6.2$ billion years. Here, lower times reflect times closer to the big bang and a greater wave front time of flight from today, $13.7$ billion years. The remaining part of the wave front, not shown, is a nearly circular and connects to two sides.

In Fig.~\ref{3waves:fig}, the scale of the individual wave fronts is the same, but the distance between the wave fronts is not to scale with the time between the wave fronts. This was done to show visually how the wave front evolves. The wave front at the bottom of the figure has just passed through the lens, and we see that it has developed a dimple in its center. As the wave front continues to evolve, it folds over itself, leading to a pair of cusp singularities. 

\begin{figure}
\begin{center}
\scalebox{1.0}{\includegraphics{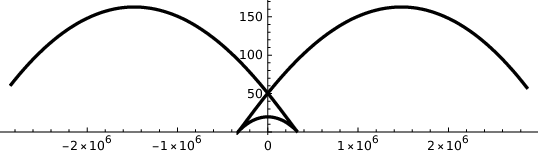}} \caption{\label{flatfront:fig}
A portion of the wavefront at 6.2 billion years. Distances along both axes are measured in light years, indicating that the depth of the cusp structure is very small compared with the breadth of this section of the wavefront. } \end{center}\end{figure}

Figure~\ref{flatfront:fig} shows the wave front at $t=6.2$ billion years with $\hat x$ and $\hat z$ axes shown in (light) years. This figure demonstrates how flat the wave front structure is in reality. The depth of the wave front, the distance between the minimum $\hat z$ value and maximum $\hat z$ value, is about 150 years while the horizontal range over which the wave front is multi-valued is about 400,000 years. The overall wave front is essentially circular, with the exception of this small singularity structure, and the overall wave front has a radius of 7 billion years. 

The full three-dimensional wave front would be obtained by rotation. Figures \ref{3waves:fig} and \ref{flatfront:fig} show the slice of the wave front in the $\hat x$-$\hat z$ plane, and rotating about the $\hat z$ axis would result in a cusp ridge in later wave fronts centered on the $\hat z$ axis. The majority of the full wave front would be spherical.

\subsection{Time Delays} \label{timedelays:sec}

For the spatial positions where multiple light rays connect the source and observer, we can consider the time of flight, or alternately, the time delay between signals. Because we are working with perturbed cosmological metrics, we assume that the light source and the observer are stationary in the spacetime. We have set up our null geodesic equations, Eqs.~\ref{wavefrontODEs:eqn}, as the past light cone of the observer. Thus, integration of the null geodesics can determine the time a light ray leaves the source to be received at the observer today.

Generally, a source with positive $x$ value has one light ray that connects to the observer through $+x$ values, and two that connect from the other side (see Fig.~\ref{4rays:fig}). The ray that travels along the right-hand side has the shortest time of flight of the three light rays. Table~\ref{times:table} shows the time differences for a source at $s=0.0002$ billion years and $s = 0.0001$ billion years. In both cases, the light from the leftmost ray arrives next and the central ray arrives last. Arriving last means that the light ray was emitted earlier in the universe's history. While the central ray has a visually shorter path, it passes most centrally through the gravitational lens, so that its longer time of flight reflects a gravitational influence on time.  

The time delays for this lens model range from $16.5$ to $35.3$ years. The lens in question represents a massive cluster of galaxies, with sources and observers very far apart. Time delays between gravitationally lensed images have not been observed for such large galaxy clusters to date, but a supernovae observed first in the rightmost image would be visible a few dozen years later in the leftmost image. When the lens is a single galaxy and the distances between the observer and source are smaller, time delays can be on the order of tens or hundreds of days. Such time delays have been observed.

\begin{table}[]
    \centering
    \begin{tabular}{c|c|c}
         $r_f \theta$ & Location & Time Delay \\
         \hline
         ~ & left &  16.5 \\ 
         0.0001 & middle & 24.7 \\
         ~ & right & - \\
         \hline
         ~ & left & 32.7  \\
         0.0002 & middle & 35.3  \\
         ~ & right & - \\
    \end{tabular}
    \caption{The time delay in years between the light from the first image and others for gravitationally lensed sources. The distance $s = r_f\theta$ is the coordinate distance from the optical axis in the source plane in billions of years. The choice of two source positions shows the two horizontal cuts of Figure~\ref{xversusb:fig} with multiple images. Light arrives first from the light ray on the far right and last from the light ray that moves through the center of the lens.  }
    \label{times:table}
\end{table}

\section{Discussion}

In this paper, we have applied the spacetime perspective to gravitational lensing in the case of cosmological solutions perturbed by a spherically symmetric matter distribution. We chose to work within the most common flat, $\Lambda$-dominated cosmological model, but the overall approach would apply to other cosmologies. We were able to demonstrate how one creates a lens mapping in Section~\ref{mapping:sec} that correctly explains the shape of observed images, and we explored the full range of lensing effects including multiple images, time delays and wave front singularities.

Historically, one of the arguments for using the thin lens approximation has been that by approximating the bending angle from the general relativity result, one can write the lens equation as an algebraic equation. Doing so substantially reduced the computational effort required to model gravitational lensing systems.

\begin{figure}
\begin{center}
\scalebox{1.0}{\includegraphics{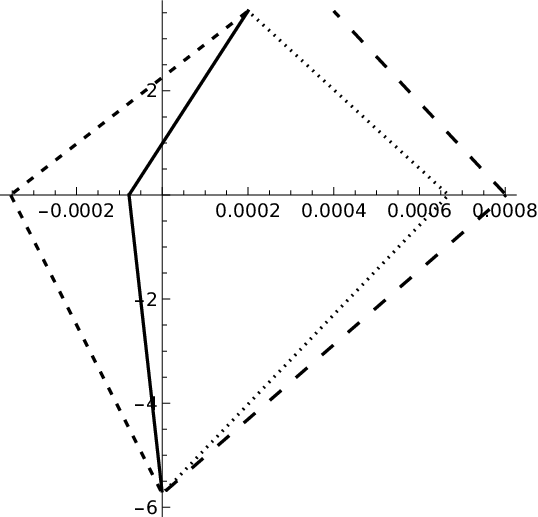}} \caption{\label{4rays:fig}
Light rays connecting an observer with sources at $s=0.0004$ billion years and $s = 0.0002$ billion years. The units of both axes are in billion years with the lens centered at the origin.} \end{center}\end{figure}

The numerical work in this paper was conducted using Wolfram Mathematica on their cloud server. Integral expressions were handled using Mathematica's NIntegrate command, and differential equations were solved using NDSolve. A typical notebook used to generate one figure, such as Fig.~\ref{4rays:fig}, ran in about five minutes or less. Wolfram Mathematica's built-in numerical integrators are modern and reasonably efficient codes, and the cloud server has strong computational capabilities. This particular work would not have been possible 20 to 25 years ago on a Mathematica installation on a typical workstation, due to the limitations of the workstation itself and Mathematica's efficiency. Our personal experience is that at this time, well-written, efficient python or C++ codes running on a strong local workstation are factors of two to ten times faster than Mathematica at similar tasks. 


Several papers have examined the accuracy of the thin lens approximation, and found conditions where the accuracy of the approximation begins to become observationally significant \cite{fk11, kf}. These papers showed, for instance, that the calculation of the gravitational lens's total mass based on measuring the Einstein ring radius using the thin lens approximation is in error on the order of $1$ to $2\%$. The errors were sensitive to the lens redshift and dependent on the truncation scheme used in mass models. 

One key argument against replacing the thin lens approximation with the methods in this paper has been based on computational effort, and this paper's results show that this argument is beginning to lose saliency, as, at least in the case of the spherically symmetric lens, one can model lensing fully, with reasonable computational run times, without the thin lens approximation. 


The spacetime approach used in this paper can be generalized to modified gravity theories. To do so, one would start with the metric that represents a physical scenario of interest, and write out the geodesic equations. This may provide an opportunity to look for differences in the gravitational lensing observables.

Expanding the techniques in this paper to more complicated and realistic lenses will further test the possibility of replacing the thin lens approximation. Future work will involve examining non-axially symmetric lensing perturbations and  lenses at multiple redshifts. With these models, one will be able to see the full range of possibilities for wave front singularities. It may also be possible to determine the non-symmetric mass distribution of a real lens based on reconstructing the source from multiple images, updating the work done for example in \citet{tyson}.




\vskip .3in

\noindent {\bf{Funding:}} SMP was supported by a summer research grant from the Massaschusetts Space Grant Consortium to participate in this work.

\noindent {\bf{Data Availability:}} All data generated or analyzed during this study are included in this published article. Mathematica codes are included in supplementary information files.

\noindent {\bf{Author Contributions:}} SMP and TPK worked together throughout the development and research phase of the paper, with SMP producing first drafts of most figures. Most writing was done by TPK and all authors reviewed the manuscript.

\end{document}